\documentclass[aps,twocolumn,showpacs,superscriptaddress,floatfix,nofootinbib]{revtex4}

\usepackage{graphicx}
\usepackage{epsfig}
\usepackage[english]{babel}

\newcommand{\identity}{\openone}
\newcommand{\be}{\begin{equation}}
\newcommand{\bea}{\begin{eqnarray}}
\newcommand{\eea}{\end{eqnarray}}
\newcommand{\ee}{\end{equation}}
\newcommand{\bra}[1]{\mbox{$\langle #1 |$}}
\newcommand{\ket}[1]{\mbox{$| #1 \rangle$}}
\newcommand{\braket}[2]{\mbox{$\langle #1  | #2 \rangle$}}
\newcommand{\proj}[1]{\mbox{$|#1\rangle \langle #1 |$}}
\def\H{{\cal H}}
\def\C{{\cal C}}
\def\O{{\cal O}}

\begin{document}

\title{Simulation of quantum dynamics with quantum optical systems}
\author{E. Jan\'e}
\affiliation{Departament d'Estructura i Constituents de la Mat\`{e}ria, Universitat de Barcelona,
Diagonal 647, E-08028 Barcelona, Spain}
\author{G. Vidal}
\affiliation{Institute for Quantum Information, California Institute for Technology, Pasadena, CA
91125 USA}
\author{W. D\"ur}
\affiliation{Sektion Physik, Ludwig-Maximilians-Universit\"at M\"unchen, Theresienstr.\ 37, D-80333
M\"unchen, Germany}
\author{P. Zoller}
\affiliation{Institut f\"ur Theoretische Physik, Universit\"at Innsbruck, A-6020 Innsbruck, Austria}
\author{J.I. Cirac}
\affiliation{Max-Planck Institut f\"ur Quantenoptik, Hans-Kopfermann Str. 1, D-85748
Garching,Germany}

\date{\today}

\begin{abstract}
We propose the use of quantum optical systems to perform universal simulation of quantum dynamics. Two specific implementations that require present technology are put forward for illustrative purposes. The first scheme consists of neutral atoms stored in optical lattices, while the second scheme consists of ions stored in an array of micro--traps. Each atom (ion) supports a two--level system, on which local unitary operations can be performed through a laser beam. A raw interaction between neighboring two--level systems is achieved by conditionally displacing the corresponding atoms (ions). Then, average Hamiltonian techniques are used to achieve evolutions in time according to a large class of Hamiltonians.
\end{abstract}


\maketitle



\section{Introduction}


Simulating quantum systems on a classical computer is known to be hard.
Consider a set of two--level quantum systems, say $N$ spin--$1/2$ particles placed at the sites of some regular lattice, that interact with each other. The number of parameters required to describe the state of these spins grows exponentially with $N$ \cite{Feynman,Jozsa}.  For instance, the state of $N=50$ spin--$1/2$ systems is specified by $2^{50} \approx 10^{15}$ numbers, whereas a $2^{50}\times 2^{50}$ matrix, i.e. with $\approx 10^{30}$ entries, needs to be exponentiated in order to compute its time evolution.
Therefore, a device processing information according to the laws of classical physics ---and in particular any present computer--- is unable to efficiently simulate the dynamics of these spins when $N$ becomes large.


This fact has severe consequences in the study of condensed matter systems. Our understanding of a large spectrum of collective quantum phenomena, including ferromagnetism, antiferromagnetism, conductor--insulator transitions, superconductivity and quantum Hall effect, is dramatically hindered by the impossibility of tracing quantum dynamics. Suppose the $N$ spins, whose $2^{N}$--dimensional Hilbert space $\H$ decomposes as
\be
\H_N = \H^{(1)} \otimes\H^{(2)} \otimes \cdots \otimes \H^{(N)},~~~~~\H^{(k)} = \C^2,
\label{eq:qubits}
\ee 
display a property that we want to study, originating, say, in a given collective effect, and we conjecture that some Hamiltonian $H_N$ acting on $\H_N$ is able to account for this effect. Since a classical computer can not efficiently simulate a quantum evolution according to $H_N$, the question whether $H_N$ successfully describes the property of interest can in general not be answered. 
And thus, for instance, the simple Hubbard model, believed to explain a wide range of electromagnetic properties of condensed matter systems --high temperature superconductivity among them--, remains unsolved after decades of study \cite{Hubbard,Sachdev}.


As a matter of fact, Feynman's initial motivation for constructing a quantum computer was the efficient simulation of quantum dynamics \cite{Feynman}. Building on the observation that a quantum system can be used to simulate another quantum system, he conjectured the existence of a universal quantum simulator UQS. 
Since then, several authors have analyzed this possibility \cite{all}. A UQS is a controlled device that, operating itself at the quantum level, efficiently reproduces the dynamics of {\em any} other many--particle quantum system that evolves according to short range interactions. Here, the assumption of some {\em degree of locality} in the interactions, implying that
the multi--particle Hamiltonian 
$H_N=\sum_{i} H_{i}$ 
is a sum of terms $H_{i}$ each one involving only a few neighboring systems, is important to achieve an efficient simulation. In most cases of interest this requirement happens to be fulfilled. Consequently, a UQS could be used to efficiently simulate the dynamics of a generic many--body quantum system and in this way function as a fundamental tool for research in quantum physics.


On the other hand, the main present motivation for building a quantum computer comes from the expected exponential gain in efficiency of certain quantum algorithms with respect to their classical counterparts. Shor's efficient factorization of large numbers is so far the most celebrated milestone of quantum computation \cite{Shor}. However, for quantum computers to overcome classical ones in tasks such as factorization, they would have to coherently operate tens of thousands of two--level systems or quantum bits (qubits). This extraordinary enterprise requires technology that may only be at reach in several decades from now. Instead, simulating the dynamics of a few tens of qubits appears as a more feasible ---and, arguably, still immensely rewarding--- task.


In this paper we propose the use of quantum optical systems to realize a UQS.  For illustrative purposes we put forward two specific schemes. First we consider the use of neutral atoms trapped in optical lattices \cite{Jaksch,Bloch}. 
 The second scheme consists of ions in an array of microtraps \cite{CZ2000}. In both cases, some raw interaction is produced between neighboring qubits by conditionally displacing the atoms (ions). Also, fast local unitary transformations are applied on the qubits, in a way that the evolution of the system effectively corresponds to that generated by some new, simulated Hamiltonian. In this way, a large class of Hamiltonians $H_N$ are produced.

The present work aims, therefore, at bridging between previous theoretical developments concerning the simulation of quantum dynamics and its experimental realization. Among the main motivations for this contribution we encounter the belief that a universal quantum simulator might be counted in the near future among the first demonstrated fundamental applications of quantum information science.


\section{Overview}\label{section:results}


In this paper we propose, through two specific realizations, the use of quantum optical systems to construct a UQS. That is, we put forward two specific schemes 
for the realization of a device that aims at simulating the dynamics of other quantum systems. 

We would like to note at the outset that from a computational point of view a UQS can be regarded as just a particular instance of a quantum computer and, conversely, any quantum computer could be used to simulate quantum systems. A simulation--oriented design, however, is likely to be decisive when it comes to actually achieving quantum simulations with existing technology \cite{uqs}. It is this search for feasibility with presently available resources what possibly best defines the general philosophy of the present work.
Instead of discussing fundamental aspects of quantum dynamics simulation, here we will be primarily concerned with presenting both concrete physical set--ups and protocols for its feasible implementation. In particular, we will describe how to achieve the simulation of a considerably large class of interesting unitary dynamics that require particularly little time and control resources when performed in the proposed physical set--ups.

A brief summary of the proposals, both having two--level systems or qubits as basic building blocks, is as follows. 

\begin{figure}
\epsfig{file=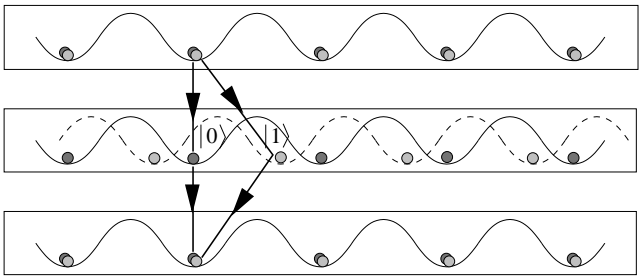,width=8.5cm} \caption{\label{fig:atoms} Atoms in a double optical lattice. An interaction between adjacent qubits is achieved by displacing one of the lattices (which traps the atoms when their internal state is $\ket{1}$, light balls) with respect to the other (that traps the atoms when their internal state is $\ket{0}$, dark balls). In this way the $\ket{1}$ component of atom $a$ approaches in space the $\ket{0}$ component of atom $a+1$, and these collide in a controlled way. Then the two components of each atom are brought back together, and local unitary transformations can be enforced by shining a laser. By a sufficiently large, relative displacement of the two lattices, also interactions between more distant qubits can be achieved. Notice that the manipulation of the qubits is homogeneous and highly parallel.} 
\end{figure}

{\bf UQS1}: Each qubit corresponds to two internal states $\ket{0}$ and $\ket{1}$ of a neutral atom. Two superposed, identical optical lattices are homogeneously filled with $N$ atoms, one atom per lattice site. Local unitary transformations are performed on the qubits by shinning the atoms with a laser beam, whereas two--qubit interactions are produced by displacing one of the optical lattices (which traps the atoms when their internal state is $\ket{1}$) with respect to the other lattice (that traps the atoms when their internal state is $\ket{0}$) [see figure \ref{fig:atoms}]. 

\begin{figure}
\epsfig{file=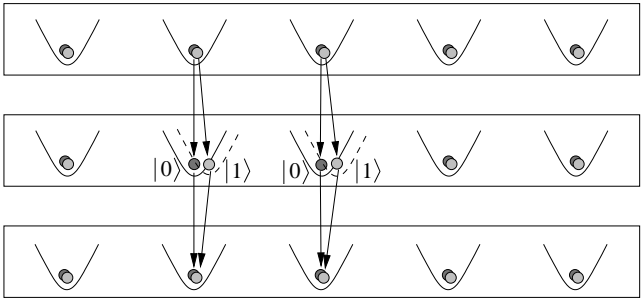,width=8.5cm} \caption{\label{fig:ions} Ions in an array of microtraps. An interaction between two qubits is achieved by conditionally pushing the corresponding ions with a force that only acts when their internal state is $\ket{1}$. The non--local evolution originates from a difference in the electrostatic potential affecting each ion, which depends simultaneously on the internal state of both ions. Local unitary operations are enforced on selected qubits by shinning a laser beam on the corresponding ions.
}

\end{figure}

{\bf UQS2}: Each qubit corresponds to two internal states $\ket{0}$ and $\ket{1}$ of an ion. $N$ ions are stored in an array of microtraps, with one ion per trap. Again, local unitary transformations are performed on the qubits by means of a laser beam. In this scheme, an interaction between two selected qubits is achieved by pushing the corresponding ions with a force that only acts if the internal state of the ion is $\ket{1}$ [see figure \ref{fig:ions}]. 

\begin{figure}
\epsfig{file=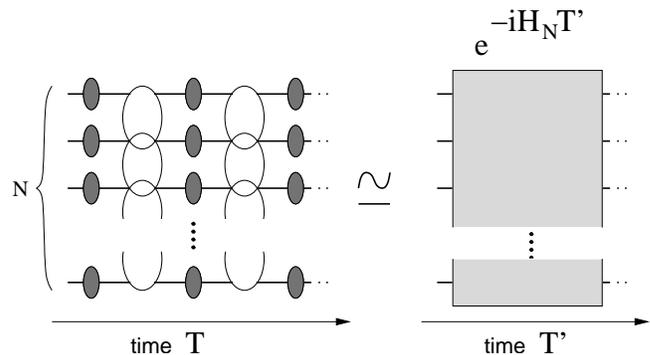,width=8.5cm}
\caption{\label{fig:sim} The simulation of a unitary evolution of $N$ qubits (horizontal lines) according to a Hamiltonian $H_N$ as in Eq. (\ref{eq:HN}) is achieved by composing individual unitary transformations on each of the qubits (dark ellipses) with short gates that decompose into two-qubit components (empty ellipses). }
\end{figure}

Schemes UQS1 and UQS2 share many features and we will take advantage of this fact by conducting a common analysis in most of the paper. Both schemes are based on qubits, as in Eq. (\ref{eq:qubits}), described by the algebra of Pauli spin--1/2 operators. A second common feature concerns the way the multi--qubit system that constitutes the simulator can be externally addressed. The manipulation is divided into two classes of external interventions that produce, respectively, {\em local unitary transformations} on each qubit and {\em two-body interactions} between qubits. Simulations are achieved by combining these two possibilities (see figure \ref{fig:sim}). Through the interaction, short \cite{short} two--qubit unitary transformations are produced. These are interspersed with fast one--qubit unitary transformations, in a way that some average Hamiltonian effectively guides the evolution of the qubits. Thus, we mimic decoupling and refocusing bang--bang techniques of nuclear magnetic resonance \cite{NMR}, although the present schemes also benefit notoriously from a more direct control of the interactions. Of considerable use will be the characterization of non--local Hamiltonian simulation recently performed in the context of quantum information \cite{qinfo,Be01,Ma02}.

The basic features of these proposals, namely the use of spin--1/2 systems and the enforcement of one--qubit and two--qubit Hamiltonian evolutions, have practical implications that are worth discussing in this introductory part.
By definition, a UQS is a device able to simulate efficiently any multi--particle quantum system with range--restricted interactions and, as mentioned above, could in principle function as a general purpose quantum computer. However, technological limitations significantly reduce the tasks that it can accomplish in practice. This observation affects also the kind of systems that a specific implementation of a UQS can effectively simulate given some experimental possibilities. In particular, the very nature of our proposals makes them best suited for simulating evolutions of systems whose building blocks are also two--level systems, and having a Hamiltonian
\be
H_N = \sum_{a} H^{(a)} + \sum_{a\not{=} b} H^{(ab)}
\label{eq:HN}
\ee
that decomposes into one--qubit terms $H^{(a)}$ and two--qubit terms $H^{(ab)}$. [Here $a$ and $b$ are indices that label the qubits]. One finds, for instance, that for this class of systems the time required to perform a simulation using our schemes is proportional to the time of simulated evolution, a very convenient fact \cite{othersystems}. Accordingly, here we will focus the discussion of simulation protocols primarily to the case of Hamiltonians of the form (\ref{eq:HN}). These include a variety of popular spin models, such as Ising interactions and both isotropic and anisotropic Heisenberg interactions \cite{Sachdev}. 


Finally, a brief comparison of the two proposals comprises the following observations. What makes scheme UQS1 most attractive is its feasibility with present technology, as supported by the results of current experiments with neutral atoms in optical lattices \cite{Bloch}. Another appealing feature is the high degree of parallelism in the manipulation of the qubits. Its major drawback is that in present experiments only homogeneous local operations can be performed ---so that all qubits experience the same local unitary transformations--- and two--qubit interactions necessarily follow a translationally invariant pattern. The degree to which each qubit can be individually addressed is an important issue, since it determines whether the scheme constitutes a universal simulator. We will eventually suggest methods to operate on each qubit independently within scheme UQS1, but for most of the paper only homogeneous manipulations are assumed, implying that only a subclass of Hamiltonians (\ref{eq:HN}) can be simulated. Scheme UQS2, in turn, is distinguished precisely by its higher degree of addressability. In this proposal each qubit can be individually manipulated, so that arbitrary (inhomogeneous) local unitary transformations can be performed. Also, the two-body interaction can be produced between arbitrary ---but not too distant--- pre--selected qubits, allowing for the simulation of any Hamiltonian of the form (\ref{eq:HN}).


The rest of the paper is distributed as follows. Section \ref{section:hamiltsim} introduces some background material on the simulation of quantum dynamics. Section \ref{section:setups} contains a discussion of the schemes UQS1 and UQS2. In section \ref{section:simulation} a series of examples are presented to illustrate how to simulate the dynamics of $N$ qubit systems in our set--ups. Finally, in section \ref{section:discussion} we discuss how to improve proposal UQS1, and we consider, in the context of studying the ground state of a multi--qubit Hamiltonian, the effect of errors and imperfections in the simulation.


\section{Simulation of quantum dynamics}\label{section:hamiltsim}

This section presents background material concerning the simulation of unitary quantum dynamics. We first describe how an effective Hamiltonian evolution can be obtained by time averaging over some other Hamiltonian evolutions. After a brief analysis of the time and control resources required in a simulation of this type, we review useful results of what is known among the quantum information community as {\em non--local Hamiltonian simulation using local unitary transformations} \cite{qinfo,Be01,Ma02}, a technique that uses fast local unitary operations on qubits to effectively modify an existing interaction between them. The reason for discussing this technique is that its requirements match the control possibilities of our two proposals.


A starting observation concerning the simulation of quantum dynamics is that if a Hamiltonian $K\equiv \sum_{i=1}^s K_i$ decomposes into terms $K_i$ acting on a small constant subspace, then by Trotter formula\ \cite{hbar}, 
\be
e^{-i K \tau} = \lim_{m\rightarrow \infty} \left( e^{-i K_1 \tau/m}e^{-i K_2 \tau/m}\cdots e^{-i K_s \tau/m} \right)^{m},
\label{eq:trotter}
\ee
we can approximate an evolution according to Hamiltonian $K$ by a series of short evolutions according to each of the pieces $K_i$. Therefore we can simulate the evolution of the $N$--qubit system of Eq. (\ref{eq:qubits}) according to Hamiltonian $H_N$ of Eq. (\ref{eq:HN}) by composing short one--qubit and two--qubit evolutions generated, respectively, by the one--qubit and two--qubit Hamiltonians $H^{(a)}$ and $H^{(ab)}$. 


In our schemes an evolution according to one--qubit Hamiltonians $H^{(a)}$ will be obtained directly by properly shinning a laser beam on the atoms or ions that host the qubits. Instead, two--qubit Hamiltonians $H^{(ab)}$ will be achieved by processing some given interaction $H_0^{(ab)}$ that is externally enforced, as we explain in the following.


Let us consider only two of the $N$ qubits, that we denote $a$ and $b$. By alternating evolutions according to some available, switchable two--qubit interaction $H^{(ab)}_0$ for times $\{t_i\}$ with local unitary transformations, one can achieve an evolution
\bea\label{evol1}
U(t)&=&\prod_{i=1}^{n} V_i\exp\left(-iH_0^{(ab)} t_i\right)V_i^\dagger\nonumber\\&=&\prod_{i=1}^{n}
\exp\left(-iV_iH_0^{(ab)} V_i^\dagger t_i\right),
\eea
where $t\equiv\sum_{i=0}^{n-1} t_i$, $V_i\equiv u_i^{(a)}\otimes v_i^{(b)}$, with $u_i$ and $v_i$ being one--qubit unitary transformations, and the products are time--ordered. We proceed by considering the case where the total time $t$ is very small as compared to the time-scale of $H_0^{(ab)}$, so that 
$-\identity \ll tH_0^{(ab)}\ll \identity$. For notational convenience, Hamiltonian $H_0^{(ab)}$ and time $t$ are set to be dimensionless, with the eigenvalues of $H_0^{(ab)}$ being of order 1. Then, for $t\ll 1$ we can approximate the exponentials in Eq. (\ref{evol1}) by the first terms in their expansion in powers of $t$, to obtain
\be\label{evol2}
U(t) \simeq \identity-it\sum_{i=1}^{n}p_iV_iH_0^{(ab)} V^\dagger_i + \O(t^2),
\ee
where $p_i\equiv t_i/t$. That is, the evolution of the two qubits corresponds, effectively, to having the Hamiltonian
\be\label{simhamilt}
H_{eff}^{(ab)}\equiv\sum_{i=1}^{n}p_iV_iH_0^{(ab)} V^\dagger_i +\O(t),
\ee
acting on them for time $t$. Then, by concatenating several short gates $U(t)$,
\be
U(t) = e^{-iH_{eff}^{(ab)}t} + \O(t^2),
\label{eq:shortgate}
\ee
we can simulate $H_{eff}^{(ab)}$ for larger times. 


Two parameters, the {\em time cost} $c$ and the {\em control complexity} $\chi$, can be used to characterize the above simulation. We consider the fast control limit, where local operations $V_i$ are performed very fast as compared to the time scale of $H_0^{(ab)}$, so that the simulation time is determined only by the use of $H_0^{(ab)}$. Suppose the aim is to simulate an evolution according to $H^{(ab)}\equiv cH_{eff}^{(ab)}$, $c>0$, for a simulated time $T'$. It follows from the linearity of Eq. (\ref{simhamilt}) that this requires using $H_0^{(ab)}$ for a time $cT'$. We define the {\em time cost} $c$ as
\be
c \equiv \frac{T}{T'},
\ee 
so that it measures the time overhead required to simulate $H^{(ab)}$ by $H_0^{(ab)}$. We remark that this measure of time resources implicitly assumes that the simulation time $T$ is proportional to the simulated time $T'$ \cite{othersystems}.
 On the other hand, a simulation that takes time $T$ is achieved by composing a number $L$ of gates $U(t)$ such that $T=Lt$. At each gate an error of order $t^2$ is introduced (see Eq. (\ref{eq:shortgate})), so that after time $T$ the total error is of order $\epsilon\equiv Lt^2$, provided $\epsilon \ll 1$. For a given error $\epsilon$ and simulated time $T'$, we find that the number of gates $U(t)$ must be $L=c^2T'^2/\epsilon$. That is, in order to obtain a constant error $\epsilon$ in a simulation for simulated time $T'$, the total number $L$ of gates $U(t)$ must grow quadratically in $T'$, while the small time step $t$ defining $U(t)$ must decrease as $t=\epsilon/(cT')$. Then, if it takes $n$ control operations $V_i$ to perform each gate $U(t)$, the total number of control operations ---per unit of simulation time $T$--- required to simulate $H^{(ab)}$ by $H_0^{(ab)}$ is given by
\be
\chi \equiv \frac{n L}{T} = \frac{ncT'}{\epsilon},
\label{eq:control}
\ee
where $\chi$ defines the {\em control complexity} (per time unit) of the simulation. 

Summarizing, a two--qubit Hamiltonian and fast local operations allow us to simulate other two--qubit Hamiltonians. The simulation time $T$ is proportional to the simulated time $T'$, and the required number $\chi$ of control operations per unit time grows linearly with $T'$ \cite{improvement}. The above analysis can now be carried out for $N$ qubits with one--qubit and two--qubit Hamiltonians. A relevant aspect is how the time cost $c$ and the control complexity scale with $N$, for this may ultimately determine whether a given simulation is feasible. We find, for instance, that whenever parallel manipulation of all $N$ qubits is possible, the time cost $c$ will not depend on $N$, whereas the control complexity will grow linearly in $N$, $\chi = ncNT'/\epsilon$.


We move now to consider the degree of independence between $u_i$ and $v_i$ in the control operation $V_i=u_i^{(a)}\otimes v_i^{(b)}$ of Eqs. (\ref{evol1})-(\ref{evol2}), which relates to the level of single--qubit addressability available in a given physical setting. The two extreme cases correspond to {\em homogeneous manipulation}, where the laser beam affects equally the two qubits, and thus $u=v$; and to {\em inhomogeneous manipulation}, where the laser beam can be sufficiently focused as to discriminate between the qubits and enforce independent evolutions on them, so that $u$ and $v$ can be arbitrary. In what follows we review the characterization, for homogeneous and inhomogeneous manipulation, of the two--qubit Hamiltonian evolutions that can be simulated by
\be
H_0^{(ab)} = \gamma \sigma_z\otimes \sigma_z,   
\label{eq:H0}
\ee
for a real $\gamma$, ---equivalently, by short gates 
\be
U \equiv e^{-i \gamma \sigma_z\otimes\sigma_z t}
\label{eq:U}
\ee
--- and by fast local operations. We quote results of \cite{Ma02,Be01}, which are time optimal, that is with minimal time cost $c$, and that happen to have a low ---possibly optimal--- control complexity $\chi$. As in Eq. (\ref{eq:H0}), we will express Hamiltonians in terms of the Pauli matrices 
\be
\sigma_x \equiv \left(\begin{array}{cc} 
0 & 1\\
1 & 0   \end{array} \right),~
\sigma_y \equiv \left(\begin{array}{cc} 
0 & -i\\
i & 0   \end{array} \right),~
\sigma_z \equiv\left(\begin{array}{cc} 
1 & 0\\
0 & -1   \end{array} \right).
\ee


{\em (i) Homogeneous manipulation} \cite{Ma02}. By using homogeneous local unitary transformations HLU as control operations $V$ in Eq. (\ref{simhamilt}), that is $V=u^{(a)}\otimes u^{(b)}$, one can simulate interaction Hamiltonians of the form
\be
H = \sum_{i,j=x,y,z}M_{ij}\sigma_i\otimes\sigma_j,~~~M_{ij}=M_{ji},
\label{eq:Hsym}
\ee
which are symmetric under exchange of the qubits. As follows from the general analysis of Ref. \cite{Ma02}, some additional restrictions on the real symmetric, $3\times 3$ matrix $M$ apply.

{\bf Result 1.} Hamiltonian $H$ can be simulated by short gates according to $H_0^{(ab)}$ in Eq. (\ref{eq:H0}) and fast HLU if and only if the sign of $\gamma$ coincides with the sign of all non--vanishing eigenvalues $\mu_i$ of $M$. The time cost $c$ of the simulation is $c = (\sum \mu_i)/\gamma$.

{\em Example 1.} Consider the ferromagnetic [antiferromagnetic] Heisenberg interaction
\be
H= J\sum_{i=x,y,z}\sigma_i\otimes\sigma_i,
\label{eq:Heisenberg}
\ee
where $J < 0$ [$J>0$]. An evolution $e^{-iHT'}$ according to this Hamiltonian can be simulated by short gates (\ref{eq:U}),
where $\gamma$ must be such that $J\gamma >0$, alternated with fast HLU, according to the collection of $n=3$ coefficients $\{p_i\}$ and local unitary $\{V_i\}$ (see Eq. (\ref{evol1})):
\bea
p_1&=&\frac{1}{3},\hspace{10pt}V_1=\identity\otimes\identity,\nonumber\\
p_2&=&\frac{1}{3},\hspace{10pt}V_2=\frac{\identity-i \sigma_x}{\sqrt{2}}\otimes\frac{\identity -i \sigma_x}{\sqrt{2}},\nonumber\\
p_3&=&\frac{1}{3},\hspace{10pt}V_3=\frac{\identity-i \sigma_y}{\sqrt{2}}\otimes\frac{\identity -i \sigma_y}{\sqrt{2}}.
\eea
The time cost for simulating (\ref{eq:Heisenberg}) with (\ref{eq:H0}) is $c=3J/\gamma$, and the control complexity $\chi$ of Eq. (\ref{eq:control}) ---or the number of control operations per unit time required in the simulation to obtain a small error $\epsilon$--- is $ncT'/\epsilon = 9JT'/(\gamma \epsilon)$.



{\em (ii) Inhomogeneous manipulation} \cite{Be01}. The possibility to perform independent operations on each of the qubits translates into the ability to simulate all possible bipartite Hamiltonians.

{\bf Result 2.} Any two--qubit interaction Hamiltonian
\be
H = \sum_{i,j = x,y,z} M_{ij} \sigma_i \otimes \sigma_j
\ee
can be achieved by alternating the short two--qubit gate $U$ of Eq. (\ref{eq:U}) with fast control operations of the form $V = u\otimes v$, that is inhomogeneous local unitary transformations LU \cite{Be01}. The time cost in time--optimal simulations is a function of the singular values $\{\mu_i \geq 0\}$ of the real, $3\times 3$ matrix $M$, and reads $c=\sum_i \mu_i/|\gamma|$. 

{\em Example 2.} An evolution according to the antisymmetric Hamiltonian
\be
H=J(\sigma_z\otimes\sigma_y-\sigma_y\otimes\sigma_z),
\ee
can be achieved by composing gates $U$ and a binary sequence of LU characterized by
\bea
p_1&=&\frac{1}{2},\hspace{10pt}V_1=\identity\otimes\frac{\identity+i \sigma_x}{\sqrt{2}},\nonumber\\
p_2&=&\frac{1}{2},\hspace{10pt}V_2=\frac{\identity-i \sigma_x}{\sqrt{2}}\otimes\identity.
\eea
The time cost of the simulation is $c=2|J|/|\gamma|$, and the control complexity can be obtained from $c$ and $n=2$.

We note here that inhomogeneous manipulation does not only have the advantage, as compared to homogeneous manipulation, that it enables the simulation of arbitrary two--qubit Hamiltonians, but also that when dealing with $N$ qubit systems, simulated interactions can depend on each couple of qubits. That is, unlike in the homogeneous case, the simulated Hamiltonian $H_N$ need not be translational invariant. A final comment concerns one qubit Hamiltonians $H^{(a)}$ of Eq. (\ref{eq:HN}). In the homogeneous case we find that $H^{(a)}$ must be the same for each qubit, whereas in the inhomogeneous case can be chosen independently for each qubit.

With these basic techniques of Hamiltonian simulation in mind, we can now proceed to the remaining sections of the paper, that explain how to simulate $N$--qubit quantum dynamics by using quantum optical systems.


\section{\label{setup}Physical set--ups}\label{section:setups}

In this section we propose and discuss two physical set--ups that can be used to simulate many--body quantum dynamics. In subsection \ref{section:optical} proposal UQS1, consisting of neutral atoms in
optical lattices, is presented. In subsection \ref{section:micro-traps} we describe a second scheme, UQS2, that consists of ions trapped in an array of micro-traps. 


\subsection{UQS1: Neutral atoms in optical lattices}\label{section:optical}


The first scheme we describe consists of neutral atoms trapped in optical lattices. Two internal states of the atoms, $\ket{0}$ and $\ket{1}$, define the qubits that are the relevant degrees of freedom for the simulated dynamics. We consider two identical, one-dimensional optical lattices with period $d$, each formed by the standing wave of two interfering laser beams that are superimposed in space \cite{3D}. The lattices are such that they trap the neutral atoms depending on their internal state, $|0\rangle$ and $|1\rangle$. We assume that the double lattice is filled homogeneously with one atom per lattice site, and that the atoms are cooled to the vibrational ground state (this can be accomplished for instance by loading the lattice with a BEC \cite{Jaksch,Bloch}). If required, the qubits can be initialized to state $\ket{0}$ through optical pumping. Their read--out can be achieved by measuring the phosphorescence of atoms after shining them with a proper laser. Depending on the degree of addressability in the scheme, only collective measurements of the qubits may be feasible.


Several proposals to perform coherent evolution of the joint state of two atoms trapped in optical lattices have been put forward \cite{Jaksch,Br99}. In \cite{So99}, it was shown how to use optical lattice systems to simulate ferromagnetism and spin squeezing. We will base our scheme on the proposals \cite{Jaksch} and \cite{So99}, although some new techniques will be introduced. Note that one may also use different techniques to achieve a controlled interaction between neighboring atoms, including recent proposals to increase possible gate fidelities of resulting phase gates by orders of magnitude \cite{Ch02}. 


In the following we will explain how the interaction between the internal degrees of freedom of different atoms can be accomplished. By adjusting the phases of the interfering laser beams, the two lattices are easily displaced with respect to each other, therefore displacing the atoms conditionally to their internal state. Performing this process adiabatically ---such that the trapped atoms remain in the motional ground state--- and choosing a relative displacement of one lattice period, we have that the $|1\rangle$ component of atom $a$ meets the $|0\rangle$ component of atom $a+1$, and the two atoms interact through controlled collisions for time $t_1$. Then the lattices are returned (adiabatically) to their initial positions. In the absence of interactions, adiabaticity requires that $|\dot{\bar{x}}(t)| \ll \nu_{\rm osc}$, where $\bar{x}$ is the relative position coordinate for the system of the two lattices, $\nu_{\rm osc} \approx a_0\omega$ is the rms velocity of the atoms in the vibrational ground state, $\omega$ is the excitation frequency and $a_0$ is the size of the ground state of the trap potential \cite{Jaksch}.

The result of the interaction process is that the $|1\rangle_a|0\rangle_{a+1}$ (as before, here $a$ labels the position of the atoms) component of the wave functions picks up a certain phase shift, which depends on the interaction time $t_1$. This is a nonlocal evolution, generated by a Hamiltonian whose interaction part reads
\be
K_1 \equiv \sum_{a} \sigma_z^{(a)} \otimes \sigma_z^{(a+1)},
\label{eq:K1}
\ee
In addition, there are other phases originating in local Hamiltonians of the form $\sum_a \sigma_z^{(a)}$. They can be removed by applying homogeneous local unitary operations on the atoms and will not be considered in what follows. Therefore, the resulting gate is of the form
\be
U_1 \equiv e^{-i\theta_1 K_1} = \prod_a \exp [-i \theta_1\sigma_z^{(a)} \otimes \sigma_z^{(a+1)}],
\label{eq:U1}
\ee
where $\theta_1$ can be adjusted by changing the interaction time $t_1$. This gate, together with homogeneous local operations on the qubits, can be used to simulate evolutions according to other {\em first neighbor} interactions.

Similarly, by displacing the lattices further, e.g. for $j$ periods with respect to each other, and allowing for interaction time $t_j$ before returning to the initial position, each of the atoms interacts with its $j^{th}$ neighbor in a controlled way. In this case a gate of the form
\be
U_j \equiv e^{-i\theta_j K_j} = \prod_a \exp [-i \theta_j\sigma_z^{(a)} \otimes \sigma_z^{(a+j)}],
\label{eq:UJ}
\ee
can be produced, where the Hamiltonian $K_j$ is given by
\be\label{opt:h0p} 
K_j= \sum_a \sigma_z^{(a)}\otimes \sigma_z^{(a+j)}, 
\ee 
and $\theta_j$ depends on the interaction time $t_j$. In this way, two-qubit interactions between $j^{th}$ neighbors can also be simulated.


Combining the above processes for different $j$ by visiting all corresponding positions one can produce a total interaction Hamiltonian $H_0$ which includes two--qubit interactions to first neighbors, second neighbors, etc.  Note that part of the time required to produce these evolutions is spent in adiabatically shifting the lattice back and forth. For $\theta_1=\pi$ in Eq. (\ref{eq:U1}), the total time $T$ required
for a complete cycle ---taking adiabaticity requirements into account--- is at the order of few microseconds \cite{Bl02}. 
This may be compared to a spontaneous emission lifetime of the single
atoms in the lattice which ranges from several seconds to many minutes.


This setup can be easily adapted also to 2D/3D arrays of optical lattices. For example, nearest neighbor interaction in a 2D square lattice is done by repeating the process for 1D in both directions of the square lattice. Note that the whole process moves all the atoms at the same time, so that the interactions that are produced are invariant under translations. Homogeneous local operations on the qubits, the most feasible class of control operations on this setting, will not break this translational symmetry. Thus, this scheme is mainly adequate to simulate translationally invariant Hamiltonians.


\subsection{UQS2: Ions in an array of micro-traps}\label{section:micro-traps}


The basic architecture of this proposal consists of $N$ ions, each one confined in an individual micro-trap \cite{CZ2000}, that we regard as a three-dimensional harmonic oscillator. Thus we consider $N$ micro--traps or independent three--dimensional harmonic oscillators, possibly generated by electric or magnetic fields \cite{DeVoe} and distributed in space according to some one-dimensional or two-dimensional pattern. Most of the present discussion will not depend on the particular geometry of the array. We only require the separation between neighboring traps to be such that a laser beam can address ions individually and that the Coulomb interaction between ions is not able to excite their vibrational state.


The relevant degrees of freedom of the simulated dynamics are two internal states of the ions, labeled as $\ket{0}$ and $\ket{1}$, that define a qubit. The initialization and readout of these qubits has been discussed in \cite{CZ95}. Since the separation between adjacent microtraps allows for individual addressing of the ions, local unitary operations on a given qubit can be performed by shinning a laser beam onto the corresponding ion, with laser frequency and phase chosen as not to affect its motion.


Next we discuss how a two--qubit interaction can be achieved \cite{CZ2000}. Suppose, for the
time being, that the ions are in the ground state of the trap\footnote{This condition will be relaxed later on.} and consider a force
$F(t)=\frac{1}{2}f(t)\hbar\omega/a_0\proj{1}$ that acts on a selected ion only when its internal
state is $\ket{1}$, producing a slight displacement in a particular direction. Here $f(t) \in [0,1]$
is a smooth function of time, $w$ is the frequency of the traps, $a_0\equiv\sqrt{\hbar/2m\omega}$ is
the typical size of the ground state of the trap and $m$ is the mass of the ion. It is important to
note that this force can be chosen to act only on a particular set of ions in the array of micro-traps
and that we can choose the direction of the displacement suffered by them.

Suppose now the force $F(t)$ acts on two ions, $A$ and $B$, which suffer a conditional displacement in the
direction $\overrightarrow{AB}$. If we ignore the rest of ions in the array, the potential for ions
$A$ and $B$ reads
\bea
V_{AB}&=&\sum_{a=A,B} \frac{m\omega^2}{2}\left[(\hat x_a -\bar x(t)\ket{1}_a\bra{1})^2
- \bar{x}(t)^2
\ket{1}_a\bra{1}\right] \nonumber \\
&+&\frac{e^2}{4\pi\epsilon_0}\frac{1}{|d+\hat x_B-\hat x_A|},\label{potential}
\eea
where $d$ is the distance between the centers of the traps, $\bar x(t)\equiv f(t)a_0$ and $\hat
x_a$ is the position operator for the ion $a=A,B$ with respect to the original equilibrium
position. Let $x_a^{(0)}$ determine the minimum of this potential when the force is not acting,
and let us replace the position operators $\hat{x}_a$ with the displacements around
$x_a^{(0)}$ and also redefine $d \rightarrow d - x_A^{(0)} +x_B^{(0)}$. We consider the case
where $|\hat x_a|\ll d$, that allows for a perturbative expansion of Coulomb potential; $\epsilon|\hat x_A
\hat x_B|/a^2_0\ll 1$, meaning that the interaction is a perturbation with respect to the traps; and $\epsilon^2\ll 1$, where $\epsilon=\frac{e^2}{4 \pi
\epsilon_0 d}\frac{2}{m \omega^2 d^2}$ is the ratio of the Coulomb energy and the energy of one ion
with respect to its neighboring trap. Under these conditions, if $f(t)$ goes from 0 to approximately
1 and then back to 0 \emph{adiabatically} (which requires $|\dot f(t)|\ll \omega$), the only effect on the ions will be the appearance of a
phase $\phi_{ij}$ depending on the internal state $\ket{i}_A\ket{j}_B$ of the ions, $i,j=0,1$. The non--local content of this evolution is given by $\phi$,
\bea
\phi &\equiv& \phi_{11} - \phi_{01} - \phi_{10} + \phi_{00}=\nonumber\\&=&
\frac{e^2}{4\pi\epsilon_0}\int_0^T \!dt \!\left[
\frac{1}{d+\bar{x}_B-\bar{x}_A}-\frac{1}{d+\bar{x}_B}-\frac{1}{d-\bar{x}_A} +\frac{1}{d} \right]
\nonumber\\&\simeq&
 -m\omega^2 \epsilon \int_0^T \! \! dt~ \bar x_A(t) \bar x_B(t),
\label{phase}
\eea
where the contribution $\phi_{ij}$ has been computed by considering the electrostatic potential at the time-dependent minimum $(\bar x_A(t), \bar x_B(t))$ of $V_{AB}$ when the internal state of the ions is $\ket{i}_A\ket{j}_B$.

The total effect of pushing the two ions back and forth is therefore to transform their internal states according to a unitary transformation $U$ that, up to local terms that can be subsequently undone by applying a proper laser beam, reads 
\be
U=e^{-i\theta \;\sigma_z^A \otimes\sigma_z^B }, ~~~ \theta \approx -\frac{e^2a_0^2}{8\pi\epsilon_0}\frac{1}{d^3}
\int_0^T dt f(t)_Af(t)_B,
\label{eq:UU}
\ee
that is, according to the interaction Hamiltonian
\be
H_0^{(AB)}=\sigma_z^A \otimes\sigma_z^B.
\ee
Then, arbitrary two--qubit Hamiltonian evolutions can be simulated between qubits $A$ and $B$ by alternating gates of the form (\ref{eq:UU}) and local operations on each qubit, as explained in the preceding section.

Note that the conditional phases $\phi_{ij}$ in (\ref{phase}) depend only on the mean position of the
ions and not on the width of their wave functions. Hence they are insensitive to the temperature,
and the ions need not be in the ground state of the traps as the force starts acting. Notice also
that both the external force $F_a(t)$ and the Coulomb potential due to the rest of ions may also
contribute to the phases $\phi_{ij}$, but they will not affect the non-local phase $\phi$.


For later reference we point out that a force homogeneously applied to {\em all} the ions naturally produces, because of the $1/d^3$ decay of $\theta$ in Eq. (\ref{eq:UU}), a gate according to the Hamiltonian
\be\label{ham:micro}
H_0=\sum_{a\neq b} \frac{1}{d_{ab}~^3}\sigma_z^{(a)} \otimes\sigma_z^{(b)},
\ee
where $d_{ab}$ denotes the distance between ions $a$ and $b$.


A final observation is that this natural $1/|a-b|^3$ decay in the strength of the interactions can be used to perform parallel processing on $N$ qubits, for large $N$. For instance, if we apply simultaneously the force $F(t)$ to, say, qubits $a$ and $a\!+\!1$ and qubits $a\!+\!10$ and $a\!+\!11$, the strength of the interaction between, say, qubit $a$ and qubit $a\!+\!10$ will be $10^{-3}$ times that between qubits $a$ and $a\!+\!1$, so that it may be neglected. In particular, this means that, for sufficiently large $N$, the time cost $c$ of simulating a Hamiltonian $H_N$ involving $N$ qubits may be made not to depend on $N$.


\section{Quantum simulator at work}\label{section:simulation}

This section presents simulations that could be carried out with the proposals UQS1 and UQS2 of section \ref{section:setups}. Only a reduced number of simple examples have been selected for this discussion, that is not meant to be exhaustive, but just intends to illustrate the broad range of applications that a universal quantum simulator would have in the study of condensed matter systems.

We recall that in both physical set--ups of section \ref{section:setups} one--qubit evolutions are achieved by shinning a laser on the systems (either atoms or ions) carrying the qubits, while two--qubit interactions are produced by conditionally displacing these systems. Then, one can use the techniques of Hamiltonian simulation of section \ref{section:hamiltsim} to achieve evolutions of $N$ qubits according to some effective Hamiltonian $H_N$.

The first three examples we discuss correspond to Hamiltonians $H_N$ that are invariant under translations. Their simulation does not require single--qubit addressability, and therefore can be performed in any of the two set--ups UQS1 and UQS2. The fourth example requires the ability, inherent in proposal UQS2, of manipulating each qubit independently. 


\subsection{Dipole--dipole interaction Hamiltonian}\label{section:dipole}

We first consider a system of $N$ spin--1/2 particles placed at the sites of some 1D or 2D regular lattice and with dipole--dipole interactions. It is described by a Hamiltonian of the form
\be\label{dd:h}
H_{D}\equiv\frac{1}{2}\sum_{a\neq b} \frac{J}{d_{ab}\,^3} \left (
\sigma_+^{(a)}\otimes\sigma_-^{(b)}+\sigma_-^{(a)}\otimes\sigma_+^{(b)}\right ),
\ee
where $\sigma_-\equiv\ket{0}\bra{1}$, $\sigma_+\equiv \ket{1}\bra{0}$, and $d_{ab}$ denotes the distance between spins at lattice sites $a$ and $b$. 

A possible interest to simulate Hamiltonian $H_{D}$ comes from the fact that, in the 2D case, with anisotropic dipole interaction --that is, with the interaction strength $J$ depending on the direction between spins-- this simple model is known to have a {\em spin--glass phase} \cite{Sachdev,spinglass}, which depends on the degree of anisotropy.

An evolution according to $H_{D}$ can be easily simulated if short evolutions according to a Hamiltonian
\be\label{dd:h0}
H_0=\frac{1}{2}\sum_{a\neq b} \frac{1}{d_{ab}\,^3} \sigma_z^{(a)}\otimes\sigma_z^{(b)}
\ee
are enforced on the $N$ qubits.
In scheme UQS1, this can be accomplished by concatenating gates $U_j$ of Eq. (\ref{eq:UJ}) for different $j$, and by tuning the interaction times $t_j$ so that coefficients $\theta_{j}$ decay with the cube of the distance between atoms. In scheme UQS2, Hamiltonian $H_0$ can be produced by simply pushing all ions simultaneously with the conditional force $F(t)$, as explained above (see Eq. (\ref{ham:micro})), since the $1/d^3$ decay appears naturally there.
In either case, short gates according to $H_0$ can be alternated with homogeneous local unitary operations on the $N$ qubits, as we did for $N=2$ in section \ref{section:hamiltsim} (see {\em example 1}). If we consider the sequence $\{p_i,V_i\}$ of weights and local control operations given by
\bea
p_1&=&\frac{1}{2},\hspace{10pt}V_1=\left( \frac{\identity-i \sigma_x}{\sqrt{2}}\right)^{\otimes N},\\
p_2&=&\frac{1}{2},\hspace{10pt}V_2=\left( \frac{\identity-i \sigma_y}{\sqrt{2}}\right)^{\otimes N},
\eea
then, we have that the average Hamiltonian reads
\bea
&&\sum_{i=1}^2 p_i V_iH_0 V_i^\dagger = \nonumber\\
&&\frac{1}{4}\sum_{a\neq b} \frac{1}{d_{ab}\,^3} \left (
\sigma_x^{(a)}\otimes\sigma_x^{(b)}+\sigma_y^{(a)}\otimes\sigma_y^{(b)}\right ),
\eea
which is equivalent to $H_{D}$, as follows from the identities $\sigma_x=(\sigma_++\sigma_-)$, $\sigma_y=-i(\sigma_+-\sigma_-)$.


\subsection{Ising and Heisenberg Hamiltonians}\label{section:ising}

We move now to discuss the simulation of the Ising Hamiltonian
\be\label{ih:ising}
H_{I}\equiv-\frac{J}{2}\sum_{\langle a,b\rangle}\sigma_z^{(a)}\otimes\sigma_z^{(b)},
\ee
as well as that of the Heisenberg Hamiltonian
\be\label{ih:heisenberg}
H_H\equiv-\frac{J}{2}\sum_{\langle a,b\rangle}\left(\sigma_x^{(a)}\otimes\sigma_x^{(b)} +\sigma_y^{(a)}\otimes\sigma_y^{(b)} +\sigma_z^{(a)}\otimes\sigma_z^{(b)}\right),
\ee
where the symbol $\langle ,\rangle$ means that the sums include only first neighboring sites in the array.

Simulating these Hamiltonians appears as a promising enterprise. For instance, the Ising model in a 2D triangular lattice and with $J<0$ can be used to study {\em frustration effects}, whereas the Heisenberg model in a 2D triangular lattice has been discussed in the context of {\em high temperature superconductivity} and {\em quantum Hall effect} \cite{Laughlin}.

The simulation of $H_I$ is particularly easy using scheme UQS1. In a 1D setting, a short gate according to $H_I$ can be achieved by displacing the lattices as explained in the previous section (recall Eq. (\ref{eq:K1})). In 2D and 3D, a lattice displacement in each spatial direction must be enforced. The achievement of only first--neighbor interactions in scheme UQS2 requires creating short gates according to the Hamiltonian $\sigma_z^{(a)}\otimes\sigma_z^{(b)}$, where $a$ and $b$ are first--neighbor qubits. This is achieved by pushing the corresponding two ions. Each couple of ions has to be pushed at a different time, in order to avoid undesired interactions between ions that are not first neighbors. However, we already mentioned that the $1/d^3$ decay in Eq. (\ref{eq:UU}) allows for parallel processing of sufficiently distant couples of ions.

The Heisenberg model $H_H$ can be simulated from the Ising model $H_I$ (see also \cite{So99}). We have already described the $N=2$ case in {\em example 1} of section \ref{section:hamiltsim}. Similarly, in the general $N$ case we can compose short gates according to $H_I$ with homogeneous local operations according to the set $\{p_i, V_i\}$ given by
\bea\label{ih:protocol}
p_1&=&\frac{1}{3},\hspace{10pt}V_1=\identity^{\otimes N}\nonumber\\
p_2&=&\frac{1}{3},\hspace{10pt}V_2=\left(\frac{\identity-i \sigma_x}{\sqrt{2}}\right)^{\otimes N},\nonumber\\
p_3&=&\frac{1}{3},\hspace{10pt}V_3=\left(\frac{\identity-i \sigma_y}{\sqrt{2}}\right)^{\otimes N}.
\eea

In these two Hamiltonians one could add an extra term of a magnetic field
\be\label{ih:b}
\tilde H=B\sum_a \sigma^{(a)}_{\vec n}
\ee
where $\vec n$ denotes a particular direction in real space. To simulate the presence of a magnetic field one has to apply a short unitary operation
\be
U=\exp \left[-i \left({\vec \sigma} \cdot {\vec n}\right) \delta t\right]^{\otimes N},
\ee
where $\vec\sigma\equiv(\sigma_x,\sigma_y,\sigma_z)$, at the end of each simulation cycle. 

Note that introducing \emph{local} terms to a given simulated Hamiltonian can always be done, irrespectively of the Hamiltonian. Whether these local terms can depend on each qubit or must be homogeneous will depend on the available degree of individual addressability in the physical scheme for simulation.


\subsection{Ising model with random coefficients and magnetic field}\label{section:random}

Let us consider a Hamiltonian of the form
\be\label{ra:ising}
H_{R}=-\frac{1}{2}\sum_{\langle a,b\rangle}J_{ab}\,\sigma_z^{(a)}\otimes\sigma_z^{(b)} + \sum_a B_a \sigma_x^{(a)},
\ee
that is, with first--neighbor Ising interaction, where coefficients $J_{ab}$ depend on the pair of qubits under consideration, and a qubit--dependent magnetic field along the $x$ direction. 

In the 1D case and with coefficients $J_{ab}$ randomly drawn from some probability distribution, this model has been used in the study {\em spin glasses} \cite{spinglass,Sachdev} and {\em percolation} \cite{Sachdev}. It has also been considered in the context of solving classical problems with {\em quantum annealing}, as compared to classical annealing methods \cite{Calz}. 

Simulating this Hamiltonian requires single qubit addressability. In the proposal UQS2, $H_{R}$ can be simulated by producing short two--qubit gates $U_{ab}$ according to $\sigma_z^{(a)}\otimes\sigma_z^{(b)}$, where qubits $a$ and $b$ are first neighbors. One possibility is to assign a frequency $\nu_{ab}$, proportional to $J_{ab}$, to each couple of qubits ($a,b$), and to perform gate $U_{ab}$ with that frequency. The magnetic field can be simulated by means of local operations, as we have described above.


\subsection{Lattice geometry, many-body Hamiltonians and universal simulations}\label{section:many}

\begin{figure}
\epsfig{file=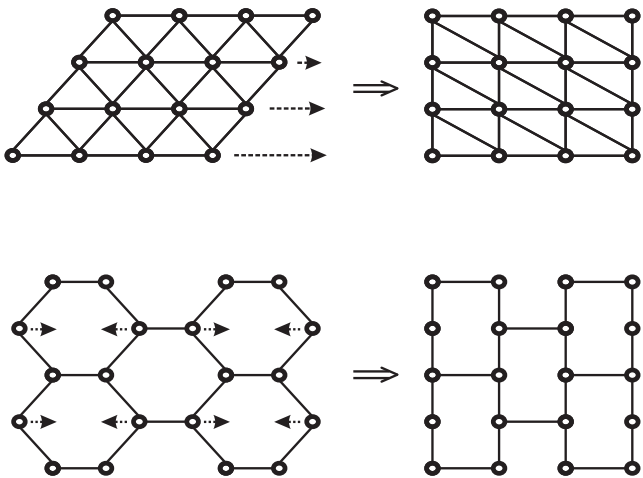,width=8.5cm}
\caption{\label{fig:geometry} Illustration how triangular or hexagonal configurations of atoms with
nearest neighbor interactions may be simulated in a rectangular array
using only nearest neighbor interactions. Solid lines symbolize
interactions between atoms.} 
\end{figure}

In practice, the physical configuration of a device designed to simulate quantum dynamics is likely to be fixed. This may imply that the lattice pattern of the simulating qubits can not be changed to adjust the desired simulation. However, different lattice patterns can be effectively achieved from a fixed one. For instance, in a two dimensional pattern, a system with nearest neighbor interaction in a triangular configuration can be obtained from a rectangular array configuration. This is achieved making the subsystems in the rectangular array interact not only with their nearest neighbor but also with two of their next-to-nearest neighbors in the same diagonal (see figure \ref{fig:geometry}).

As explained in the preceding sections, proposals UQS1 and UQS2 are best suited to simulating $N$--qubit Hamiltonians with one--qubit and two--qubit terms. In these cases, the simulation time is proportional to the simulated time and the control complexity $\chi$ is remarkably low. For $n>2$, $n$--body Hamiltonian terms can also be simulated in these schemes by exploiting the following identity
\be\label{mb:prot}
e^{-i H_1 \theta}e^{-i H_2\theta}e^{i H_1\theta}e^{i H_2\theta}=e^{[H_1,H_2]\theta^2+\mathcal{O}(\theta^3)},
\ee
where $[A,B]=AB-BA$ is the commutator of operators $A$ and $B$. A short gate according to a three body Hamiltonian $H$ that can be expressed as the commutator of two two--body Hamiltonians $H_1$ and $H_2$, that is $H=-i[H_1,H_2]$, is achieved by concatenating the four short two--body gates on the lhs of Eq. (\ref{mb:prot}). As an example, with $H_1\equiv\identity\otimes\sigma_z\otimes\sigma_z$ and $H_2\equiv\sigma_x\otimes\sigma_x\otimes\identity$, one can obtain an evolution given by Hamiltonian $\sigma_x\otimes\sigma_y\otimes\sigma_z$.
A detailed analysis shows, however, that the time $T$ required to simulate a three--body interaction by two--body interactions grows quadratically in the simulated time $T'$. Arbitrary $n$--body interactions can be simulated in this way, with similar polynomial scaling of the simulation time.


Also $d$--level systems, $d>2$, can be simulated by a qubit--based scheme for quantum dynamics simulation. For this purpose one needs to group the qubits into subsets of $l$ neighboring qubits, where $l$ is the smallest integer greater than or equal to $\log_2 d$. Single--particle Hamiltonian terms for each $d$--level systems will correspond to $l$--qubit dynamics. Interactions between $n$ $d$--level systems will require the simulation of $(n \times l)$--qubit interactions.


In this way, a set of qubits with a switchable two--qubit interaction and the ability to enforce independent local unitaries on each qubit is sufficient to perform universal simulation of quantum dynamics, as is the case of our second proposal. Instead, when all the control evolutions are homogeneous, as in scheme UQS1, only Hamiltonians that are invariant under translations can be simulated. In the next section we will indicate possible ways to break this translational symmetry, thereby endowing scheme UQS1 with the capability of performing universal simulation of quantum dynamics.


\section{Discussion}
\label{section:discussion}


As argued in section I, a general purpose quantum computer may only be feasible in several decades from now. Instead, a quantum simulator ---a device designed with the specific purpose of simulating other multi--particle quantum systems--- could render the simulation of quantum dynamics feasible in a much closer future, as present experimental results \cite{Bloch} encourage us to believe. This paper intended to provide a connection between previous theoretical considerations \cite{Feynman,all} and the actual realization of a device for simulating quantum dynamics. With this purpose we have put forward two specific experimental set--ups and practical mechanisms to attain a large class of Hamiltonian evolutions.


Certainly, much future work is required to complement the brief analysis presented here. In this last section we would like to describe some of the aspects that should be further investigated, and initiate their analysis. In particular, we shall examine some of the limitations of scheme UQS1 and put forward possible solutions. The effect of several kinds of errors is also investigated numerically in the context of a specific application of a quantum simulator of quantum dynamics: the preparation of the unknown ground state of a simulated Hamiltonian $H_N$.

\subsection{Limitations and imperfections}


{\em Addressability.} One of the most significant limitations of proposal UQS1 is the lack of independent addressability of individual qubits. In present day experiments performed with optical lattices, the distance between trapping sites, $d$, is smaller than the best achievable focusing width of lasers beams, $w$. One typically encounters $d/w\sim 10^{-1}-10^{-2}$, so that several tens of neighboring qubits are affected by the same manipulation. Although homogeneous Hamiltonians, that is, those invariant under translations, can be still simulated with this scheme, it would be most desirable to find mechanisms to achieve independent qubit manipulation. In what follows we mention three possibilities:


($i$) A spatially dependent magnetic field can be applied to shift the energy levels that define the qubits. In this way, the original energy difference between the two levels, $E_0$, becomes spatially dependent, $E(\vec r)$, and this allows to address the atoms independently. If we shine the system with a laser with frequency $\omega_L$, it will only be resonant with the atoms in the lattice site $\vec r_L$, where $\omega_L=E(\vec r_L)$ \cite{Bl02}.


($ii$) The second possibility is to use the spatial dependence of the intensity of the lasers. Consider a laser beam centered at $r_0$ (the positions refer to the plane perpendicular to the laser beam) with an intensity shape proportional to $f(|\vec r_0-\vec r|)$, where $f(0)=1$. If this laser is driving the transition $\ket{1}\leftrightarrow\ket{0}$ of an atom sitting at $\vec r$, the unitary operation acting on the atom can be written as $U=\exp(-i t \nu_0 f(|\vec r_0-\vec r|) \sigma_x )$, where $\nu_0$ is the coupling constant when the laser is centered at the atom position. Suppose we have a chain of atoms at positions $\{\vec r_j\}$, and we want to perform the unitary operation $V_a=\exp(-i \tau \sigma_x)$ on atom $a$ leaving the rest unaffected. The procedure is to aim one laser to each atom $j$ performing the unitary transformation $U_j=\exp(-i t_j \nu_0 \sigma_x)$ for $j\neq a$ and $U_a=\exp(i t_a \nu_0 \sigma_x)$ on atom $a$. Since the laser beams overlap, the total operation on atom $j$ is 
\be
V_j=\exp(-i \nu_0\sum_{k=1}^N t_k f(|r_j-r_k|)(-1)^{\delta_{ka}}\sigma_x).
\ee
Now, the condition to be applied to perform the desired unitary operation on atom $a$ is $V_j=\identity$ for $j\neq a$ and $V_a=\exp(-i \tau \sigma_x)$ and it is obtained by solving the following system of linear equations,
\be
\tau_j=\nu_0\sum_{k=1}^N t_k f(|r_j-r_k|)(-1)^{\delta_{ka}}, \hspace{10pt}\tau_j=\tau\delta_{ja}.
\ee


($iii$) The last option is to increase the distance between the trapping sites $d$. This can be accomplished by changing the geometry ---namely the relative angle--- of the two interfering lasers that form the trap.


{\em Imperfections.} So far in this work we have assumed the ideal situation where no errors occur during the manipulation of the physical set--up that constitute the quantum simulator. In practice, however, one has to deal with deviations from this ideal regime, due to several coexisting sources of errors: uncontrolled interactions of the atoms (ions) with the environment will cause {\em decoherence} in the $N$-qubit space; {\em timing errors} in {\em laser pulses} will translate into unitary gates (either local or non--local) that differ from the intended ones; the {\em adiabaticity} condition in the performance of short non--local gates in scheme UQS1 (UQS2) may not be sufficiently fulfilled, resulting in the excitation of the motional degrees of freedom of the atoms (ions); $\O(t^2)$ corrections in the expansion of Eq. (\ref{eq:shortgate}); and others. 


It is beyond the scope and possibilities of the present paper to analyze the effect of such imperfections. We have nonetheless performed a number of numerical simulations involving up to 9 qubits, in the context of a particular application of a quantum simulator, namely the study of the ground state and first excited states of a multi--qubit Hamiltonian. The results, described in the next subsection, suggest the scheme is considerably robust against several kinds of errors.


It is also worth mentioning here that decoupling techniques \cite{decoupling} can be applied in our schemes to overcome some of the difficulties we have mentioned above. In particular, undesired local phases appearing during the simulation of Hamiltonians can be eliminated using fast \emph{homogeneous} local unitary transformations. This procedure may also be useful to eliminate inhomogeneous local phases originating in (uncontrolled) variations of the magnetic field while using method ($i$) above to enhance the addressability in proposal UQS1. The procedure is based on the basic observation that a one--qubit evolution according to Hamiltonian $H_L \equiv \sigma_z$ can be undone by a similar evolution according to Hamiltonian 
\be
VH_L V^{\dagger} = -\sigma_z = -H_L,
\ee
 where $\{V\equiv i\sigma_x,V^{\dagger}\}$ denote fast local unitary operations, whereas the two--qubit Hamiltonian $H_{NL} \equiv \sigma_z\otimes\sigma_z$ is left unchanged when the two qubits are similarly rotated, 
\be
(V\otimes V)H_{NL}(V\otimes V)^{\dagger}=H_{NL}.
\ee
Therefore, in proposals UQS1 and UQS2, where the conditional pushing of atoms or ions produces a short gate $U$ according to a Hamiltonian of the form 
\be
\sum_{a} \gamma_a H_L^{(a)} + \sum_{ab} \gamma_{ab} H_{NL}^{(ab)}, 
\ee
a gate according to only two--qubit Hamiltonians $H_{NL}^{(ab)}$ can be accomplished by performing $U$, a rotation of each qubit according to $V^{\dagger}$, another gate $U$ and a final rotation of each qubit by $V$. 

 
\subsection{Application: studying the ground state of a multi-particle Hamiltonian} \label{subsection:groundstate}

The following, final discussion considers a particular application of a universal simulator of quantum dynamics, namely the study of the ground state of a simulated Hamiltonian. This is a subject of remarkable interest in, for instance, the context of {\em quantum phase transitions} \cite{Sachdev}. Once the ground state of a Hamiltonian has been attained in the $N$--qubit device, one may attempt to experimentally determine its properties, such as correlation functions, as well as study the propagation of externally induced perturbations.


The simulation of quantum dynamics by using quantum systems does not reduce to achieving a controlled evolution according to some desired Hamiltonian. Also mechanisms to prepare some convenient initial state for the $N$--qubits as well as to make measurements that give information on the simulated dynamics are needed. In section \ref{section:setups} we already mentioned how to prepare all the qubits in the initial state $\ket{0}^{\otimes N}$, as well as how to perform measurements on the qubits. However, in order to study the low temperature dynamics of a given Hamiltonian $H_N$, the initial state $\ket{0}^{\otimes N}$ must be transformed into some other (possibly mixed) state of $H_N$ with much lower energy.


This can be achieved, for instance, by coupling the $N$ qubits to a thermal bath at low temperature, following the line of thought of \cite{terh}, where conditions on the coupling Hamiltonians between the system and the bath were studied, so that their joint evolution leads to {\em thermal equilibration} of the system. In the present case, however, with a sequence of gates being continuously applied to the $N$-qubits, the results of \cite{terh} may not apply straightforwardly, and a careful study of which kind of coupling Hamiltonians lead to thermal equilibrium might be required.


\begin{figure}
\epsfig{file=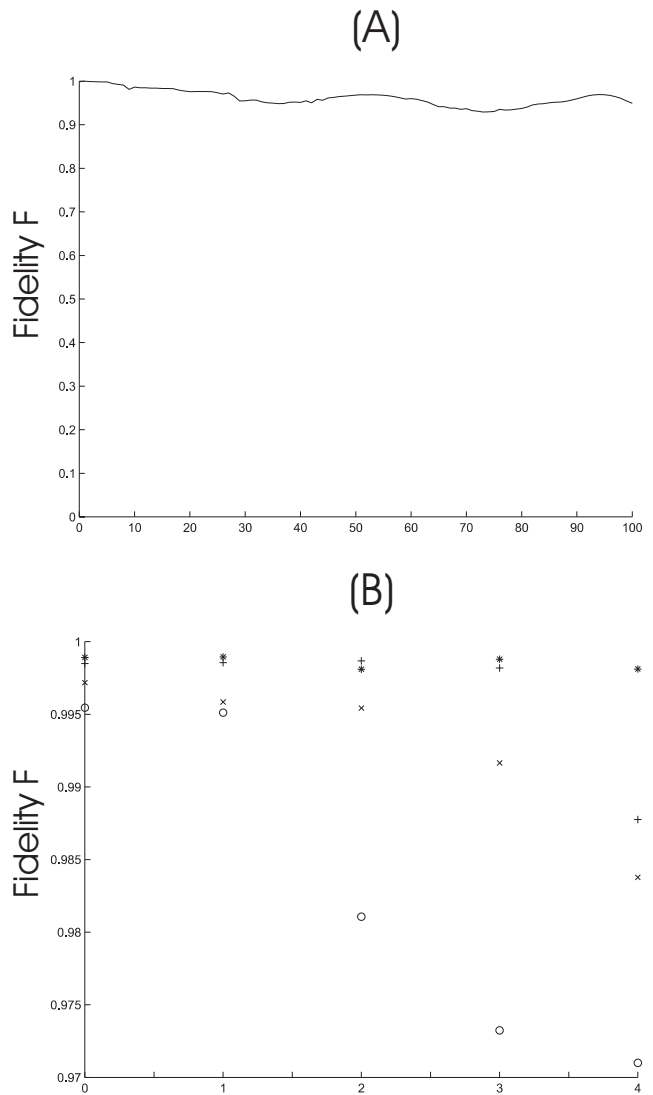,width=8.5cm}
\caption{\label{fig:hamsim}
Obtainment of the ground state of a string of $N=7$ qubits under the presence of dipole--dipole interactions by starting in the ground state of the initial Hamiltonian $H_N^0 \equiv \sum_a \sigma_z^{(a)}\otimes \sigma_z^{(a+1)}$ and adiabatically changing the simulated Hamiltonian to the dipole--dipole Hamiltonian $H_D$ (see section \ref{section:simulation}). The simulation includes timing errors in the local and two--qubit gates, and also errors due to the first order approximation in the expansion of Eq. (\ref{evol2}). In Fig. \ref{fig:hamsim}(a) we assume timing errors in the local laser pulses of 1 percent and in the interaction times of 0.5 percent. $100$ time steps are performed in the adiabatic introduction of $H_D$, with a value of $\theta_1$ in Eq. (\ref{eq:U1}) of $0.1$ at each step. The curve corresponds to the fidelity $|\braket{\Psi(t)}{\Psi_{sim}(t)}|^2$ of the state $\ket{\Psi_{sim}(t)}$ obtained through simulation with respect to the instantaneous ground state $\ket{\Psi(t)}$ of $H(t)$. Fig. \ref{fig:hamsim}(b) shows the fidelity of the final state with respect to the ground state of $H_D$ for errors of 0, 1, 2, 3 and 4 percent in both, interaction time and timing of local laser pulses. In this case we have $\theta_1=0.025$,  and the adiabatic change of the Hamiltonian is performed by using 100 (o), 250 (x), 500 (+) or 1500 (*) time steps.}
\end{figure}

An alternative approach is to make an adiabatic introduction of the Hamiltonian $H_N$ from some local Hamiltonian $H_N^{0}$ whose ground state $\ket{\Psi_0}$ can be prepared. Suppose that the aim is to bring the $N$ qubits, initially in state $\ket{\Psi_0}$, into the ground state $\ket{\Psi}$ of $H_N$. This can be achieved by simulating a time--dependent Hamiltonian $H(t)$ of the form
\be
H(t) \equiv k(t)H_N^0+(1-k(t))H_N,
\label{eq:Ht}
\ee
where $k(t)$ is a monotonic function that smoothly goes from 1 to 0 during a time $T_{\mathrm{sim}}$. The time $T_{\mathrm{sim}}$ necessary to obtain the ground state $\ket{\Psi}$ depends on energy difference $E_{e}(t)$ between the ground state and the first excited state of $H'(t)$, and must typically fulfill $T_{\mathrm{sim}} \gtrsim \max_{\{t\}} E_e(t)^{-1}$. Strictly speaking, the ground state $\ket{\Psi}$ of $H_N$ will only be achieved in the limit of large $T_{sim}$. However, for a sufficiently large time $T_{sim}$ compatible with the above condition, the final state of the system will be a mixed state $\rho$ with a large projection on the ground state $\ket{\Psi}$ of $H_N$ and some other contributions corresponding to the first excited states. Thus, one can view $\rho$ as a thermal state with very low temperature.

\begin{figure}
\epsfig{file=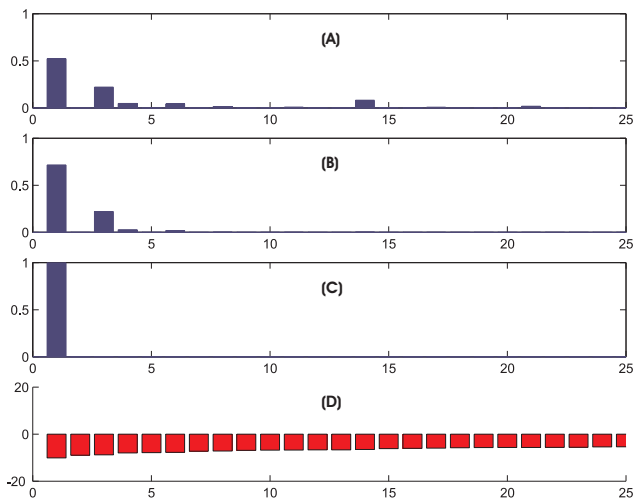,width=8.5cm}
\caption{\label{fig:ground} Obtainment of the ground state of Dipole-Dipole interaction Hamiltonian $H_D$ (Eq. (\ref{dd:h})) of a string of $N=9$ atoms in a 1D optical lattice with error parameters of one percent for two--qubit gates and timing
of laser pulses, and $\theta_1 = 0.025$ in Eq. (\ref{eq:U1}). The adiabatic
change of the Hamiltonian $\sum_a \sigma_x^{(a)}\otimes\sigma_x^{(a+1)}$ into $H_D$ is performed by means of 50 (a), 100 (b),
500 (c) time steps. The histogram shows the weight of the state obtained
via simulation in the (eventually degenerated) eigenspaces $P_j$
corresponding to the sorted eigenvalues of $H_D$, the lowest of which are plotted in
Fig. 6(d). One observes that only the low--energy eigenstates are populated, and that the probability of being in the ground state converges toward unity if the adiabaticity requirement (corresponding to the change of H(t) in Eq. (\ref{eq:Ht})) is fulfilled, that is, if more time steps are performed.}
\end{figure}

{\em Numerical simulations} of this quantum simulation, for up to 9 qubits and including several sources of errors, seem to indicate that, indeed, the final mixed state $\rho$ of the $N$--qubit system is essentially the ground state of $H_N$, together with small contributions of the first excited states (see figures (\ref{fig:hamsim})-(\ref{fig:ground})). That is, the final state of one such quantum simulations could be used to explore the low temperature properties of the simulated Hamiltonian $H_N$.

\begin{acknowledgments}

We thank I. Bloch, H.-J. Briegel, L.-M. Duan, A. A. Garriga and Ll. Masanes, for useful comments and discussions. 
This work was supported by the 
European Community under project EQUIP (contract IST-1999-11053) and grant HPMF-CT-2001-01209 (W.D., Marie Curie Fellowship) and the ESF; 
by the Institute for Quantum Information GmbH, Austria; 
by the Deutsche Forschungsgemeinschaft through ''Schwerpunktsprogramm Quanteninformationsverarbeitung'', Germany; 
by the MEC (AP99), Spain; and 
by the National Science Foundation of USA, under grant EIA--0086038.

\end{acknowledgments}



\begin{thebibliography}{99}
\bibitem{uqs} Any valid scheme for universal quantum computation can be used, in principle, to simulate quantum systems, and the schemes for a UQS presented in this paper are nothing but particular realizations of a quantum computer. In practice, however, experimental limitations drastically interfere with the ideal performance of any of these devices. Imperfect manipulations and finite decoherent times introduce errors at a rate far above the estimated threshold for achieving fault--tolerant controlled quantum evolutions. In this situation, a quantum computer that operates by, say, performing CNOT gates and one--qubit rotations is notoriously less succesful at tracing the dynamics of an array of interacting qubits than a device especially designed for this purpose.
\bibitem{othersystems}
Multi--qubit systems with $n$--body interactions, $n>2$, $d$--level systems with $d>2$ and more exotic systems such as fermions can also be simulated using our schemes. In this cases we encounter, however, that the time required to perform the simulation does no longer grow linearly in the simulated time, but according to some higher order polynomial.
\bibitem{hbar} Through the paper we use the convention $\hbar = 1$.
\bibitem{3D} For simplicity, we describe only a 1D optical lattice. For 2D [3D] lattices, one uses four (six) interfering laser beams, one pair for each dimension.
\end{thebibliography}
\end{document}